\documentclass[aps,showpacs,preprint]{revtex4}

\usepackage{graphicx}
\usepackage{epstopdf}
\usepackage{dcolumn}

\begin{document}
\title{Partial conservation of seniority and nuclear isomerism}
\author{P.~Van Isacker}
\affiliation{Grand Acc\'el\'erateur National d'Ions Lourds,
CEA/DSM--CNRS/IN2P3, BP~55027, F-14076 Caen Cedex 5, France}
\author{S.~Heinze},
\affiliation{Institute of Nuclear Physics, University of Cologne,
Z\"ulpicherstrasse 77, 50937 Cologne, Germany}
\date{\today}

\begin{abstract}
We point out the possibility
of the {\em partial} conservation of the seniority quantum number
when most eigenstates are mixed in seniority but some remain pure.
This situation occurs in nuclei for the $g_{9/2}$ and $h_{9/2}$ shells
where it is at the origin of the existence
of seniority isomers in the ruthenium and palladium isotopes.
It also occurs for $f$ bosons.
\end{abstract}
\pacs{03.65.Fd, 21.60.Cs, 21.60.Fw}
\maketitle

The seniority quantum number
was introduced by Racah
for the classification of electrons in an $l^n$ configuration
where it appears as a label
additional to the total orbital angular momentum $L$,
the total spin $S$,
and the total angular momentum $J$~\cite{Racah43}.
About ten years later it was adopted in nuclear physics
for the $jj$-coupling classification of nucleons
in a single $j$ shell~\cite{Racah52,Flowers52}.
These studies made clear the intuitive interpretation of seniority:
it refers to the number of particles that are not in pairs
coupled to angular momentum $J=0$.
In nuclear physics
the concept of seniority has proven extremely useful,
especially in semi-magic nuclei
where only one type of nucleon (neutron or proton) is active
and where seniority turns out to be conserved
to a good approximation.

Since the papers of Racah and Flowers appeared,
a wealth of further results has been obtained
and it is by now well understood
what are the necessary and sufficient conditions
for an interaction to conserve seniority
(see chapters~19 and 20 of Ref.~\cite{Talmi93}).
To give a precise definition of these conditions,
we introduce the following notations.
We consider a system of $n$ particles with angular momentum $j$
where for the sake of generality
$j$ can be integer for bosons or half-integer for fermions.
A rotationally invariant two-body interaction $\hat V$
between the particles
is specified by its $\lfloor j+1\rfloor$ matrix elements
$\nu_\lambda\equiv\langle j^2;\lambda
|\hat V|j^2;\lambda\rangle$
(where $\lfloor x\rfloor$ is the largest integer
smaller than or equal to $x$).
The notation $|j^2;\lambda\rangle$
implies a normalized two-particle state
with total angular momentum $\lambda$
which can take the values $\lambda=0,2,\dots,2p$,
where $2p=2j$ for bosons and $2p=2j-1$ for fermions.
The interaction can then be written as
$\hat V=\sum_\lambda\nu_\lambda\hat V_\lambda$
where $\hat V_\lambda$ is the operator defined via
$\langle j^2;\lambda'|\hat V_\lambda|j^2;\lambda''\rangle=
\delta_{\lambda\lambda'}\delta_{\lambda\lambda''}$.

With the above conventions the necessary and sufficient conditions
for the conservation of seniority can be written as
\begin{equation}
\sum_\lambda a^\lambda_{jI}\nu_\lambda=0,
\qquad
I=2,4,\dots,2p,
\label{e_consen}
\end{equation}
with
\begin{eqnarray}
&&\frac{a^\lambda_{jI}}{\sqrt{2\lambda+1}}=
\delta_{\lambda I}+2\sqrt{(2\lambda+1)(2I+1)}
\left\{\begin{array}{ccc}j&j&\lambda\\ j&j&I\end{array}\right\}
\nonumber\\
&&-
\left[\frac{16(2\lambda+1)(2I+1)}
{(2j+1)(2j+\sigma)(2j+2+\sigma)(2j+1+2\sigma)}\right]^{1/2},
\nonumber
\end{eqnarray}
where the symbol between curly brackets
is a Racah coefficient
and $\sigma\equiv(-)^{2j}$ is $+1$ for bosons
and $-1$ for fermions.
These conditions have been derived previously in a variety of ways
mostly for fermions~\cite{Talmi93,Rowe01,Rosensteel03}.
Although Eq.~(\ref{e_consen}) determines all constraints
on the matrix elements $\nu_\lambda$
by varying $I$ between 2 and $2p$,
it does not tell us how many of those are independent.
This number turns out to be
$\lfloor j/3\rfloor$ for bosons and $\lfloor(2j-3)/6\rfloor$ for fermions,
the number of independent seniority $v=3$ states~\cite{Ginocchio93}.

Conservation of seniority does not, however, imply solvability.
In general,
even if an interaction satisfies the conditions~(\ref{e_consen})
and conserves seniority,
that does not imply that closed algebraic expressions
can be given for its eigenenergies and eigenfunctions.
As regards its characterization from the point of view of symmetries,
seniority can be viewed as a {\em partial} dynamical symmetry.
It is important to clarify first
what exactly is meant by a partial dynamical symmetry
which is an enlargement of the concept of dynamical symmetry
as defined, {\it e.g.}, in chapter~11 of Ref.~\cite{Iachello06}.The idea is to relax the conditions of {\em complete} solvability
and this can be done in essentially two different ways:\begin{enumerate}
\item
{\it Some of the eigenstates keep all of the quantum numbers.}In this case the properties of solvability, good quantum numbers,and symmetry-dictated structure are fulfilled exactly,
but only by a subset of eigenstates~\cite{Alhassid92,Leviatan96}.
\item
{\it All eigenstates keep some of the quantum numbers.}
In this case none of the eigenstates is solvable,yet some quantum numbers (of the conserved symmetries)
are retained.In general, this type of partial dynamical symmetry arises
if the hamiltonian preserves some of the quantum numbers
in a dynamical-symmetry classification
while breaking others~\cite{Leviatan86,Isacker99}.\end{enumerate}
Combinations of 1 and 2 are possible as well,
for example, if some of the eigenstates
keep some of the quantum numbers~\cite{Leviatan02}.

How do seniority-conserving interactions fit in this classification?
If the conditions~(\ref{e_consen}) are satisfied
by an interaction $\hat V$,
all its eigenstates carry the seniority quantum number $v$
and, consequently, the second type
of partial dynamical symmetry applies.
The eigenstates are not solvable in general
but must be obtained from a numerical calculation.
Nevertheless, {\em some} eigenstates are completely solvable
for a general seniority-conserving interaction.
This was shown by Rowe and Rosensteel~\cite{Rowe01,Rosensteel03}
who derived closed, albeit complex, expressions
for the energies of some multiplicity-free
({\it i.e.}, unique for a given particle number $n$,
angular momentum $J$ and seniority $v$) 
$n$-particle states in a $j=9/2$ shell.
This implies a partial dynamical symmetry of the first kind.
So, we conclude that seniority-conserving interactions
in general satisfy the second type of partial dynamical symmetry
but with the added feature that some multiplicity-free states
are completely solvable.

In this Letter we carry the analysis
of seniority conservation one step further
and we investigate the problem
whether it is possible
to construct interactions that in general do not conserve seniority
but which have {\em some} eigenstates with good seniority.
We recover an example of this phenomenon
which was pointed out earlier for the $j=9/2$ shell
by Escuderos and Zamick~\cite{Escuderos06} and by Zamick~\cite{Zamick07},
and we find that it also occurs for $f$ bosons.

To shed light on this problem of partial seniority conservation,
we analyze the four-particle case.
The motivation for doing so is that
the conditions~(\ref{e_consen}) can be derived
from the analysis of the three-particle case~\cite{Talmi93}.
We might thus expect possible additional features
to appear for four particles
which will indeed be confirmed by the analysis below.

A four-particle state can be written as $|j^2(R)j^2(R');J\rangle$
where two particles are first coupled to angular momentum $R$,
the next two particles to $R'$
and the intermediate angular momenta $R$ and $R'$ to total $J$.
This state is not (anti-)symmetric in all four particles
and can be made so
by applying the (anti-)symmetry operator $\hat P$,
\begin{eqnarray}
&&|j^4[II']J\rangle
\propto{\hat P}|j^2(I)j^2(I');J\rangle
\nonumber\\
&&=\sum_{RR'}\;
[j^2(R)j^2(R');J|\}j^4[II']J]\;
|j^2(R)j^2(R');J\rangle,
\nonumber
\end{eqnarray}
where $[j^2(R)j^2(R');J|\}j^4[II']J]$
is a four-to-two-particle coefficient of fractional parentage (CFP).
The square brackets $[II']$ label the four-particle state
and indicate that it has been obtained
after (anti-)symmetrization of $|j^2(I)j^2(I');J\rangle$.
The label $[II']$ defines an overcomplete, non-orthogonal basis,
that is, not all $|j^4[II']J\rangle$ states with $I,I'=0,2,\dots,2p$
are independent.
It is implicitly assumed
that $I$ and $I'$ as well as $R$ and $R'$ are even.

The four-to-two-particle CFPs are known in closed form
in terms of $9j$ symbols and, furthermore,
the overlaps $\langle j^4[II']J|j^4[LL']J\rangle$ and
the matrix elements $\langle j^4[II']J|\hat V_\lambda|j^4[LL']J\rangle$
can be expressed in terms of them.
The expressions are rather cumbersome
and are not given here
but it is accepted that the overlaps and matrix elements
are known as algebraic expressions
of the intermediate and final angular momenta.

We assume in the following that $J\neq0$,
corresponding to four-particle states with seniority $v=2$ or $v=4$.
By definition a seniority $v=2$ four-particle state is
\begin{equation}
|j^4,v=2,J\rangle=|j^4[0J]J\rangle.
\label{e_v2}
\end{equation}
A seniority $v=4$ state is
constructed from $|j^4[II']J\rangle$ with $I,I'\neq0$
and it is orthogonal to the state~(\ref{e_v2}).
It can thus be written as
\begin{eqnarray}
\lefteqn{|j^4[II'],v=4,J\rangle}
\nonumber\\
&&\qquad=|j^4[II']J\rangle-
\langle j^4[II']J|j^4[0J]J\rangle|j^4[0J]J\rangle.
\label{e_v4}
\end{eqnarray}
If there is more than one $v=4$ state for a given $J$,
the indices $[II']$ serve as an additional label.
Seniority conservation of the interaction $\hat V$ implies
\begin{equation}
\langle j^4,v=2,J|\hat V|j^4[II'],v=4,J\rangle=0
\end{equation}
or
\begin{equation}
\frac{\langle j^4[0J]J|\hat V|j^4[II']J\rangle}
{\langle j^4[0J]J|\hat V|j^4[0J]J\rangle}=
\langle j^4[0J]J|j^4[II']J\rangle.
\nonumber
\end{equation}
Insertion of the values for the four-to-two-particle CFPs
yields the conditions~(\ref{e_consen}).

We now turn our attention to the problem of partial seniority conservation
and derive the conditions for an interaction $\hat V$
to have {\em some} four-particle eigenstates with good seniority.
Note that there are a number of  `trivial' examples of this.
For example, if the total angular momentum $J$ is odd,
a four-particle state cannot be of seniority $v=0$ or $v=2$
and must necessarily have seniority $v=4$.
Also, for $J>2p$ the four-particle state must be of seniority $v=4$.
These trivial cases are not of interest here.
Instead, we study the situation
where both $v=2$ and $v=4$ occur for the same $J$
and where a general interaction $\hat V$
mixes the $v=2$ state with a subset of the $v=4$ states
but not with all.
A general seniority $v=4$ state
is specified by the coefficients $\eta_{II'}$ in the expansion
\begin{equation}
|j^4\{\eta_{II'}\},v=4,J\rangle=
\sum_{II'}\eta_{II'}|j^4[II'],v=4,J\rangle,
\label{e_v4gen}
\end{equation}
where the sum is over $q$ linearly independent combinations $[II']$
(with $I\neq0$ and $I'\neq0$),
as many as there are independent $v=4$ states. 
Let us now focus on bosons with $j\leq5$ or fermions with $j\leq13/2$.
In these cases Eq.~(\ref{e_consen}) yields only one condition
and a general interaction
can be written as a {\em single} component $\hat V_\lambda$
plus an interaction that conserves seniority.
Consequently, if the condition of partial seniority conservation
is satisfied by a single $\lambda$ component,
it will be valid for an arbitrary interaction.
The fact that~(\ref{e_v4gen}) is an eigenstate of $\hat V_\lambda$
and that this interaction does not mix it with the $v=2$ state is expressed by 
\begin{eqnarray}
\lefteqn{\sum_{II'}\eta_{II'}
\langle j^4[LL'],v=4,J|\hat V_\lambda|j^4[II'],v=4,J\rangle}
\nonumber\\
&&=E_\lambda
\sum_{II'}\eta_{II'}
\langle j^4[LL'],v=4,J|j^4[II'],v=4,J\rangle,
\nonumber\\
\lefteqn{\sum_{II'}\eta_{II'}
\langle j^4,v=2,J|\hat V_\lambda|j^4[II'],v=4,J\rangle=0.}
\label{e_conpsenab}
\end{eqnarray}
There are $q+1$ unknowns:
the $q$ coefficients $\eta_{II'}$ and the energy $E_\lambda$.
Equations~(\ref{e_conpsenab}) are also $q+1$ in number
and together with a normalization condition on the coefficients $\eta_{II'}$
they define an overcomplete set of equations in $\{\eta_{II'},E_\lambda\}$
not satisfied in general but possibly for special values of $j$ and $J$.
Furthermore, according to the preceding  discussion,
if these equations are satisfied for one $\lambda$,
they must be valid for all $\lambda$
and in each case the solution yields $E_\lambda$,
the eigenvalue of $\hat V_\lambda$.
A symbolic solution of the Eqs.~(\ref{e_conpsenab})
(for general $j$ and $J$) is difficult to obtain
but, using the closed expressions for the overlaps and matrix elements,
it is straightforward to find solutions for given $j$ and $J$.
In particular, a solution of the overcomplete set of equations
is found for $j=9/2$ and $J=4,6$.
We thus confirm the finding of Refs.~\cite{Escuderos06,Zamick07}
who noted the existence of these two states
that have the distinctive property of having exact seniority $v=4$
for {\em any} interaction $\hat V$ (barring accidental degeneracies).
Solution of the Eqs.~(\ref{e_conpsenab}) for $j=9/2$ and $J=4,6$
allows the explicit construction of the two states:
\begin{widetext}
\begin{eqnarray}
|(9/2)^4,v=4,J=4\rangle&=&
\sqrt{\frac{2363}{1570}}|(9/2)^4[22],v=4,J=4\rangle
-\sqrt{\frac{65}{5338}}|(9/2)^4[24],v=4,J=4\rangle,
\nonumber\\
|(9/2)^4,v=4,J=6\rangle&=&
\sqrt{\frac{1620896}{635341}}|(9/2)^4[24],v=4,J=6\rangle
-\sqrt{\frac{5725}{635341}}|(9/2)^4[44],v=4,J=6\rangle.
\nonumber\\
\label{e_j9wave}
\end{eqnarray}
\end{widetext}
These states are normalized
but expressed in terms of basis states that are not orthonormal.
In addition, the solutions $E_\lambda$ can be used
to derive the following energy expressions:
\begin{eqnarray}
E[(9/2)^4,v=4,J=4]&=&
\frac{68}{33}\nu_2+\nu_4+\frac{13}{15}\nu_6+\frac{114}{55}\nu_8,
\nonumber\\
E[(9/2)^4,v=4,J=6]&=&
\frac{19}{11}\nu_2+\frac{12}{13}\nu_4+\nu_6+\frac{336}{143}\nu_8.
\nonumber
\end{eqnarray}
These expressions give the absolute energies of the two states
and are valid for an {\em arbitrary} interaction among $j=9/2$ fermions.
The states are completely solvable,
independent of whether the interaction conserves seniority or not.
Their excitation energies $E_{\rm x}$
are not known in closed form, however,
since the $J^\pi=0^+$ ground state
is not solvable for a general interaction.
In contrast, a generally valid result
is the difference between the excitation energies,
which can be written as
\begin{eqnarray}
&&E_{\rm x}[(9/2)^4,v=4,J=6]-E_{\rm x}[(9/2)^4,v=4,J=4]
\nonumber\\
&&\quad=-\frac{1}{3}E_{\rm x}[(9/2)^2,J=2]
-\frac{1}{13}E_{\rm x}[(9/2)^2,J=4]
\nonumber\\
&&\quad\phantom{=}+\frac{2}{15}E_{\rm x}[(9/2)^2,J=6]
+\frac{18}{65}E_{\rm x}[(9/2)^2,J=8],
\nonumber
\end{eqnarray}
associating the excitation energies
of the $J=4$ and 6,
seniority $v=4$ states in the four-particle system
with those of the $J=2$, 4, 6 and 8,
seniority $v=2$ states in the two-particle system.

\begin{figure}
\includegraphics[width=8.5cm]{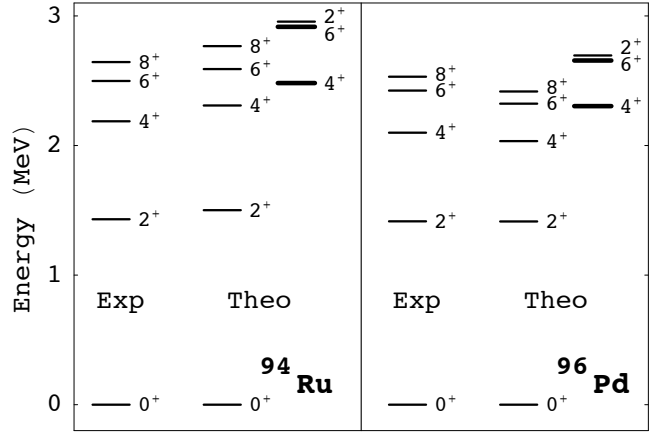}
\caption{
\label{energsen}
Experimental and calculated energy spectra
of $^{94}$Ru and $^{96}$Pd.
The $^{94}$Ru and $^{96}$Pd spectra
are calculated with $g_{9/2}$ interactions
derived from $^{92}$Mo and $^{98}$Cd, respectively,
which are seniority breaking.
All levels up to 3~MeV are shown.
The two solvable $v=4$ states are indicated in thick lines.}
\end{figure}
Another interaction-independent result that can be derived
concerns transition matrix elements.
For example, the electric quadrupole transition
between the two states~(\ref{e_j9wave})
is characterized by the $B$(E2) value
\begin{eqnarray}
&&B({\rm E}2;(9/2)^4,v=4,J=6\rightarrow(9/2)^4,v=4,J=4)
\nonumber\\
&&\qquad
=\frac{209475}{176468}B({\rm E}2;(9/2)^2,J=2\rightarrow(9/2)^2,J=0).
\nonumber
\end{eqnarray}
This again defines a parameter-independent relation
between a property of the two- and four-particle systems.

\begin{figure}
\includegraphics[width=8cm]{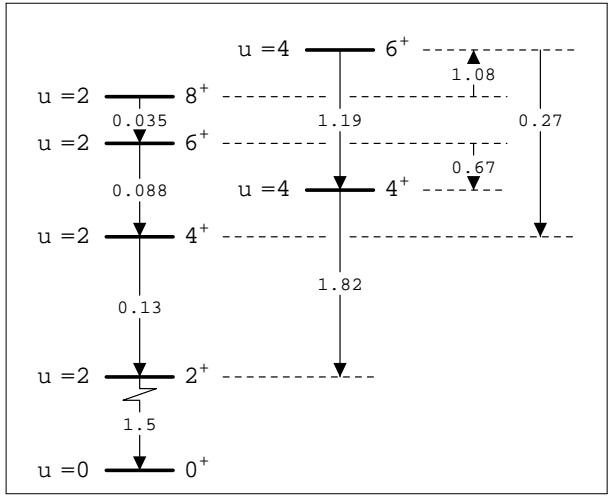}
\caption{
\label{be2sen}
E2 decay in the $(9/2)^4$ system
as obtained with a seniority-conserving interaction.
The numbers between the levels
denote $B$(E2) values expressed
in units of $B({\rm E2};2^+_1\rightarrow0^+_1)$
of the two-particle system.}
\end{figure}
There are several nuclear regions
with valence neutrons or protons predominantly confined
to an orbit with $j=9/2$,
which can be the $1g_{9/2}$ or $1h_{9/2}$ shell.
Of particular interest are the nuclei $^{94}$Ru ($Z=44$)
and $^{96}$Pd ($Z=46$)
which have four proton particles or holes in the $1g_{9/2}$ shell
and a closed $N=50$ configuration for the neutrons.
The yrast $J=2,4,6,8$ states in both isotopes
can, to a good approximation, be classified by seniority $v=2$~\cite{Grawe02}.
For any reasonable interaction
the solvable $J=4,6$ states are only a few hundreds of keV
above the $v=2$ states with the same $J$.
This is illustrated in Fig.~\ref{energsen}
which shows the observed yrast states in $^{94}$Ru and $^{96}$Pd
and compares them with the levels
calculated with two different interactions
derived from $^{92}$Mo and $^{98}$Cd, respectively.
For a constant interaction the $^{94}$Ru and $^{96}$Pd spectra
(four particles and four holes in the $g_{9/2}$ shell) are identical.
The difference between the calculated spectra in Fig.~\ref{energsen}
gives an idea of the uncertainty on the energy
which might be of use in the experimental search
for the $J^\pi=4_2^+,6_2^+$ states~\cite{Mills07}.

Partial seniority conservation sheds also
some new light on the existence of isomers
as observed in this region~\cite{Jaklevic69}.
Figure~\ref{be2sen} illustrates
the E2 decay in the $(9/2)^4$ system
as obtained with a seniority-conserving interaction.
It displays a pattern of very small $B$(E2) values
between $v=2$ states
which is typical of the seniority classification
in nuclei near mid shell ($n\approx j+1/2$)
and which is at the basis of the explanation
of seniority isomers~\cite{Grawe02}.
The decay of the two solvable $J=4,6$ states is qualitatively different,
with $B$(E2) values that are an order of magnitude larger.
The results derived here imply that,
in spite of being close in energy,
the two solvable $v=4$ states {\em do not mix}
with the $v=2$ states,
even for an interaction that does not conserve seniority.
Within a $(9/2)^4$ approximation,
the pattern shown in Fig.~\ref{be2sen} is stable
since any breaking of the seniority quantum number
of the yrast $J=4,6$ states
can occur only through mixing with the other $v=4$ levels
which lie more than 1~MeV higher.
Furthermore, the $v=4$ components in the yrast states
can be probed by detecting the M1 decay
out of the solvable $v=4$ states
since the M1 operator cannot connect components
with different seniority.

A search for solutions of Eqs.~(\ref{e_conpsenab})
did not reveal other cases of partial seniority conservation
in fermionic systems with other $j$ and/or $J$.
However, numerical studies~\cite{Heinzeun}
have shown its existence in bosonic systems,
in particular for $f$ bosons,
where we have been able to find analytic energy expressions
for several boson numbers, again valid for a general interaction.
These findings suggest
that the mechanism of partial seniority conservation with an arbitrary interaction
occurs in systems that are `only just' not entirely solvable
({\it i.e.}, $j=9/2$ for fermions and $j=3$ for bosons).
This will the subject of future investigations.

We wish to thank Larry Zamick, Alex Brown, Ami Leviatan, and Igal Talmi
for illuminating discussions.


\end{document}